\documentclass[twocolumn,showpacs,amsmath,amssymb,floatfix]{revtex4}
\usepackage{graphicx}
\begin{document}

\title{Quantum Oscillations in the Underdoped Cuprate
YBa$_2$Cu$_4$O$_8$}

\author{E. A. Yelland$^{1,*}$, J. Singleton$^2$, C. H. Mielke$^2$,
N. Harrison$^2$, F. F. Balakirev$^2$, B. Dabrowski$^3$, J. R.
Cooper$^4$} \affiliation{$^1$H.H. Wills Physics Laboratory,
Tyndall Avenue,
Bristol BS8 1TL, UK.\\
$^2$National High Magnetic Field Laboratory, MS-E536, Los Alamos
National Laboratory, Los Alamos NM 87545, USA.\\
$^3$Department of Physics, Northern Illinois University, De Kalb,
IL 60115, USA.\\
$^4$Cavendish Laboratory, J.J. Thomson Avenue, Cambridge CB3 0HE,
UK.}

\date{\today}

\begin{abstract}
We report the observation of quantum oscillations in the
underdoped cuprate superconductor YBa$_2$Cu$_4$O$_8$ using a
tunnel-diode oscillator technique in pulsed magnetic fields up to
85\,T. There is a clear signal, periodic in inverse field, with
frequency $660\pm15$\,T and possible evidence for the presence  of
two components of slightly different frequency. The quasiparticle
mass is $m^*=3.0\pm0.3 m_e$.  In conjunction with the results of
Doiron-Leyraud \textit{et al.} for YBa$_2$Cu$_3$O$_{6.5}$
\cite{DoironLeyraud07}, the present measurements suggest that
Fermi surface pockets are a general feature of underdoped copper
oxide planes and  provide information about the doping dependence
of the Fermi surface.
\end{abstract}

 \maketitle

The mechanism for high-temperature superconductivity in the
layered copper oxide superconductors has remained elusive for more
than twenty years. At the heart of the problem is the evolution of
the ground state from a Mott-Hubbard insulator to a superconductor
as the number of doped holes $p$ per planar CuO$_2$ unit is
increased. In particular, there is no agreement as to how the
underdoped region should be described.  The recent observation of
quantum oscillations in the oxygen-ordered ortho-II phase of
YBa$_2$Cu$_3$O$_{6.5}$ (O-II Y123) with $T_\mathrm{c}=57.5$\,K,
and $p=0.1$ \cite{DoironLeyraud07} shows that it has charged
quasiparticles and a well-defined Fermi surface (FS) at low
temperatures.  In this Letter we report observations of quantum
oscillations in the stoichiometric double-chain cuprate
YBa$_2$Cu$_4$O$_8$ (Y124) with $T_\mathrm{c}=80$\,K, and $p=0.125$
\cite{dopingnote} at fields up to 85\,T, suggesting that they
could be a general feature of underdoped cuprates. Our data for
Y124 show that the FS pockets expand as $p$ is increased and give
a higher quasiparticle mass $m^*$ than for O-II Y123.

The Y124 crystal was grown from flux in a ZrO$_2$ crucible under
600 bar of O$_2$ at 1100$^\circ$C. Other crystals from the same
batch were of high quality with a residual Cu-O chain resistivity
$\leq 1$\,$\mu\Omega$\,cm, and a low-$T$ thermal conductivity peak
$\kappa_b(20$\,K$)=120$\,Wm$^{-1}$K$^{-1}$ \cite{EAYthesis}.
Pulsed magnetic fields up to 85\,T were provided by the Los Alamos
85\,T multi-shot magnet \cite{Harrison07}. Measurements were made
using a tunnel-diode oscillator (TDO) technique
\cite{Coffey00,Mielke01} in which two small counter-wound coils
form the inductance of a resonant circuit. The crystal was cut
into four pieces, each measuring up to
$0.35\times0.25\times0.12$\,mm$^3$, which were stacked with their
$c$-axis directions aligned within $2^\circ$ of each other, and
placed in one coil with the $c$-axis parallel to $B$ and the axis
of the coil. The resonant frequency, in our case $47\,$MHz, can
depend on both the skin-depth (or, in the superconducting state,
the penetration depth) and the differential magnetic
susceptibility of the sample \cite{resistivitynote}. The sample
and coil were immersed in $^{3}$He liquid or $^3$He exchange gas,
temperatures ($T$) being measured with a Cernox thermometer 5\,mm
away from the sample.

Fig.~1(a) shows the TDO frequency $f$ versus $B$ at $T=0.53$\,K. At $B\approx 45$\,T,
$f$ falls substantially indicating an increase in the penetration of the rf field as
the superconductivity is suppressed. In the expanded view of the raw data taken during
the falling part of the pulse, oscillations are visible for fields $B>55$\,T. The solid
lines in Fig.~1(b) show the second derivative $\mathrm{d}^2 f/\mathrm{d}B^2$ of data
taken at 1.6\,K and reveal a clear oscillatory signal. The frequency and phase are
nearly the same during the rising (36\,T to 85\,T in 5\,ms) and falling (85\,T to 36\,T
in 10\,ms) parts of the pulse, ruling out spurious heating and electrical interference
effects.

The standard Lifshitz-Kosevich (LK) form for the oscillatory magnetization is $M\propto
B^{1/2} R_D R_T \sin(2\pi F/B+\phi)$ \cite{ShoenbergBook}, where $\phi$ is a phase, and
in conventional metals the oscillation frequency $F$ is related to a zero-field
extremal FS cross-section $A$ by the Onsager relation $F=(\hbar/2\pi e) A$
\cite{ShoenbergBook}; the scattering and temperature damping factors are respectively
$R_D=\exp(-\pi \hbar k_F/e \ell B)$ where $k_F$ is the Fermi wave vector, $\ell$ is the
mean free path and $R_T=(14.69 m^* T/m_e B)/\sinh(14.69 m^* T/m_e B)$. The dashed line
in Fig.~1(b) shows $\mathrm{d}^3 M/\mathrm{d}B^3$ \cite{resistivitynote} calculated
from the LK formula with $F=660$\,T, $\phi=\pi/2$, $m^*=3.0 m_e$, $\ell=400$\,\AA~  and
a suitable scale factor. Note that this estimate of $\ell$ assumes pure de Haas-van Alphen oscillations; any Shubnikov-de Haas component would imply a higher value. The model describes the data well, the decrease in amplitude with
$B$ arising from the weak $B$ dependence of $R_D$ at $B\sim 70$\,T and the factor of
$1/B^{6}$ in the third derivative. Note however that the non-monotonic $B$-dependence of
the oscillation amplitude at $T=0.53$\,K and $T=1.6$\,K in Fig.~2 (but not in Fig.~1b due to the $1/B^6$ factor) may be signs of beating between two close frequencies.

Fig.~2 shows $\Delta f$, the TDO frequency minus a smooth monotonic background \cite{backgdnote}, versus $1/B$, at various temperatures. The oscillations are periodic in $1/B$ as expected from quantized cyclotron orbits of Fermi-liquid-like quasiparticles. They are also damped rapidly at higher $T$, consistent with thermal smearing of the FS. The agreement in frequency and phase with Fig.~1(b) shows that the oscillations are not an artefact of background subtraction.

In Ref.~\onlinecite{DoironLeyraud07} it was pointed out that no
hole pockets are present in a  band calculation for O-II Y123
\cite{Bascones05}. However small hole pockets of mainly chain
character can be formed by allowing small shifts of the Fermi
level $\Delta E_\mathrm{F}\approx 25$\,meV \cite{Carrington07}.
Our observation of quantum oscillations in Y124, for which
calculations find no small pockets near $E_\mathrm{F}$
\cite{Carrington07}, suggests that the FS pockets are likely to be
a general feature of the copper oxide planes of underdoped
cuprates.

 \begin{figure}
\begin{center}
\includegraphics{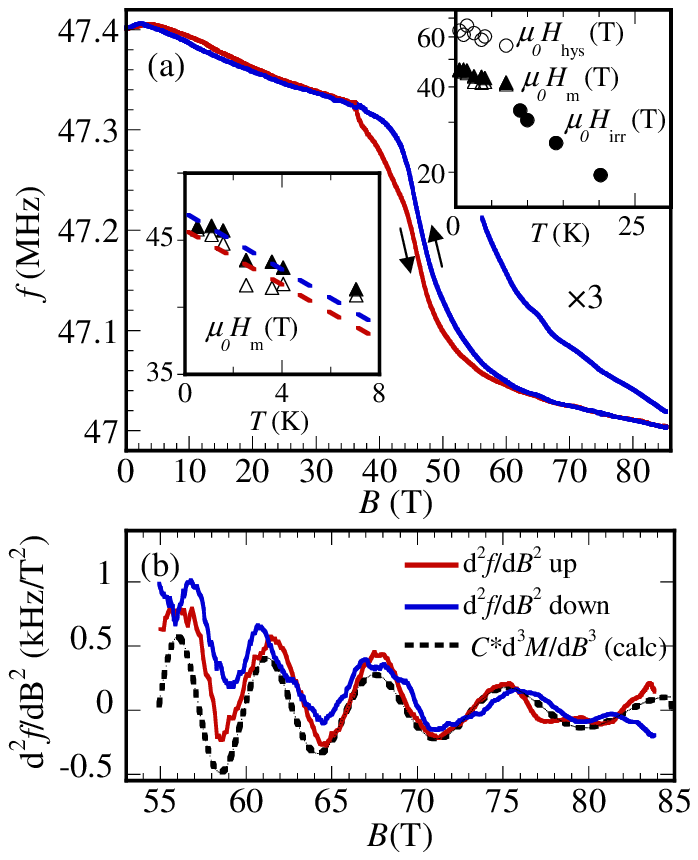}
\end{center}
\caption{\label{fig:fig1} (color online) (a) Resonant frequency
$f$ of the TDO versus magnetic field $B$ recorded during an 85\,T
pulse at $T=0.53$\,K.   Upper inset: $\mu_0 H_\mathrm{hys}$
($\circ$) and $\mu_0 H_\mathrm{m}$ (rising - $\triangle$, falling
- $\blacktriangle$) from this work and $\mu_0 H_\mathrm{irr}$
($\bullet$) from Ref.~\onlinecite{Cooperunpub}. Lower inset shows
expanded view of the $\mu_0 H_\mathrm{m}$ data at low $T$ . The
dashed lines are guides to the eye. (b) Second derivative $\mathrm{d}^2 f/\mathrm{d}B^2$ of 1.6\,K data (solid lines).  The dashed line is given by LK theory assuming $\Delta f\propto \mathrm{d}M/\mathrm{d}B$ \cite{resistivitynote}, and using a suitable scale factor.}

\end{figure}

\begin{figure}
\begin{center}
\includegraphics*[width=0.75\linewidth,clip]{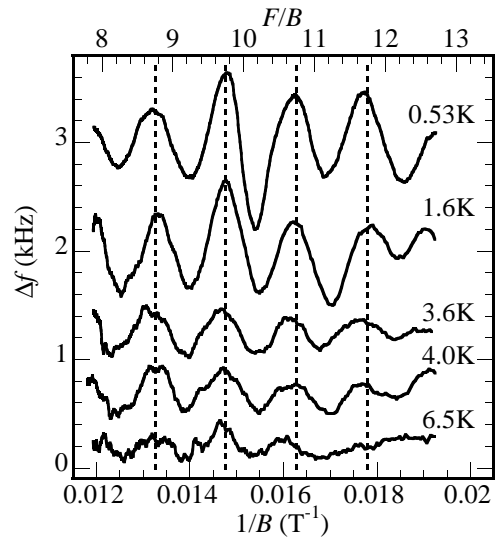}
\end{center}
\caption{ \label{fig:fig2} Changes in resonant frequency $\Delta
f$ of the tunnel-diode oscillator circuit versus $1/B$ recorded
during 85\,T pulses at various temperatures. A smooth monotonic
background has been subtracted \cite{backgdnote}. The dotted lines
are equally spaced in $1/B$. The oscillatory signal is periodic in
$1/B$ with frequency $F=660\pm15$\,T.}
\end{figure}
The insets to Fig.~1 show the $T$-dependence of  $\mu_0 H_\mathrm{m}$, the field of the
well-defined peak in $|\mathrm{d}f/\mathrm{d}B|$ and $\mu_0 H_\mathrm{hys}$ where the
hysteresis between the rising- and falling-field curves ceases to be detectable.
$H_\mathrm{m}$ is where vortex pinning becomes weak enough for the rf field to
penetrate further than the London penetration depth (but still less than the normal
state skin depth). In cuprate superconductors, vortex pinning becomes very small above
an irreversibility field $H_\mathrm{irr}$, which is usually much less than the
estimated upper critical field $H_{c2}$, although these two fields may converge as
$T\rightarrow0$.  Our values of $H_\mathrm{m}$ are similar to $H_\mathrm{irr}$
determined previously using torque magnetometry \cite{Cooperunpub} on another crystal
from the same batch (upper inset) and very recently from the resistivity of other Y124
crystals \cite{Bangura07}. Somewhat unexpectedly we find that for the present crystal
$\mu_0 H_\mathrm{hys}$ is 20\,T larger than $\mu_0H_\mathrm{m}$. The sudden onset of
hysteresis at $B=36$\,T occurs when the insert magnet is energized. This and
experiments in a faster-sweeping 65\,T magnet show that the  hysteresis increases with
$|\mathrm{d}B/\mathrm{d}t|$. The lower inset shows our values of $\mu_0
H_\mathrm{m}(T)$ on a larger scale \cite{Tnote}.

The frequency determined from LK fits to the data in
Fig.~2 and the peak positions in the Fast Fourier transform (FFT)
spectra shown later in Fig.~3(a) both give $F=660\pm 15$~T. This
corresponds to a FS pocket of only 2.4\% of the Brillouin zone
(BZ) area $A_\mathrm{BZ}$ ($\frac{\hbar}{2\pi
e}A_\mathrm{BZ}=27.9$\,kT for Y124 \cite{GinsbergBook}). If we
ascribe the oscillations to four hole-pockets as suggested for O-II Y123
\cite{DoironLeyraud07}, the hole density
$p_\mathrm{QO}=0.195\pm0.005$ compared to $p=0.125\pm0.005$
estimated from the $a$-axis thermopower \cite{Zhou96,Tallon95}.
For O-II Y123, the corresponding values of $p_\mathrm{QO}=0.152\pm
0.006$ and $p=0.1$ \cite{DoironLeyraud07} also differ by a factor
1.5. If antiferromagnetism \cite{Chen07,Harrison07b} or other order doubles the unit cell, there would be only four half-pockets in the reduced BZ and $p_\mathrm{QO}$ would be a
factor 2 smaller. The same reduction in $p_\mathrm{QO}$ is
given by earlier calculations using the $t$-$J$ model
\cite{Trugman90}. In both cases there is a discrepancy between $p$ and $p_\mathrm{QO}$ but
 this is not an issue if both electron and hole pockets are present \cite{LeBoeuf07}.

FFTs are shown in Fig.~3(a) for all temperatures measured, for the rising and falling
parts of the pulse. The amplitudes of the peak at $660$\,T were fitted to
$R_\mathrm{T}$ giving $m^*=3.0\pm0.3 m_e$ as shown in Fig.~3(b). Fig.~3(c) shows the
results of a separate LK analysis of the amplitudes of a $\sin$ curve fitted to the
data between 67 and 77~T, giving $m^*=3.10\pm 0.3 m_e$. Our best value is $m^*=3.0\pm
0.3 m_e$.

\begin{figure}
\begin{center}
\includegraphics{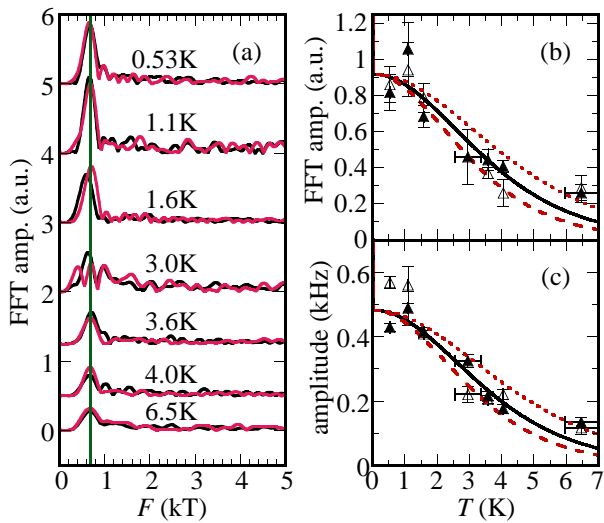}
\end{center}
\caption{ \label{fig:fig3} (a) (color online) Fast Fourier transforms in $1/B$ of
$\Delta f(B)$ over the range $60<B<85$\,T. The red (black) lines show data for the
rising (falling) part of the pulse. A single peak is present in the FFTs with a
frequency $F=660\pm15$\,T. An extra, less reproducible peak near 200 T has been removed from some of the FFTs  by subtracting a slowly varying background. (b) FFT amplitude versus $T$. Open (closed) symbols show
rising (falling) field data. The solid line shows the LK damping factor $R_T$ with
best-fit value $m^*=3.0\pm0.3m_e$. (c) amplitude of oscillatory function of the form
$\sin(2\pi F/B+\phi)$ with $F=660$\,T fitted to $\Delta f(B)$ in the range
$67<B<77$\,T. The best-fit $R_T$ curve, shown by a solid line, has $m^*=3.1\pm0.3m_e$.
Dashed lines show the LK formula for $m^*=2.5m_e$ and $m^*=3.5m_e$. }
\end{figure}

\begin{figure}
\begin{center}
\includegraphics[width=7.0cm,keepaspectratio=true]{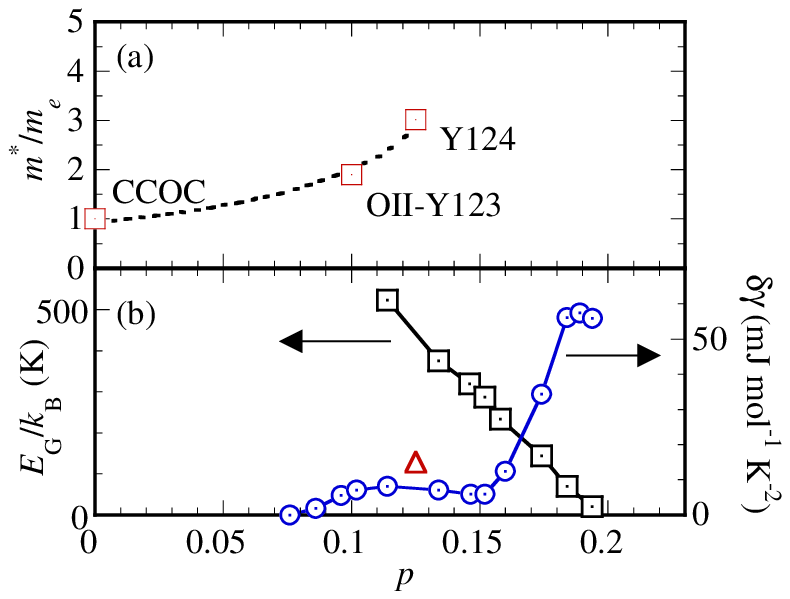}
\end{center}\caption{ \label{fig:fig4} (color online) (a) $m^*/m_e$ values
($\boxdot$) for Y124 (this work), O-II Y123 \cite{DoironLeyraud07}
and Ca$_2$CuO$_2$Cl$_2$ (CCOC), the latter from the dispersion of
Cu-O orbital states well below the chemical potential measured by
ARPES \cite{Ronning98}. CCOC is a parent Mott insulator
with $p=0$ and no FS. The dashed line is a guide to the eye. (b) Heat
capacity anomaly $\delta \gamma$ at $T_\mathrm{c}$ for various
YBa$_2$Cu$_3$O$_{6+x}$ samples ($\odot$) and the pseudogap energy
$E_\mathrm{G}$ ($\boxdot$) extracted from the same heat capacity
data using a triangular gap model \cite{Loram01}. $\delta \gamma$
at $T_\mathrm{c}$ is also shown($\triangle$) for Y124
\cite{Loramunpub}. }
\end{figure}

Fig.~4a shows the overall variation of $m^*/m_e$ with $p$ that is obtained by combining the present
result with that of Ref.~\onlinecite{DoironLeyraud07}. The value $m^*/m_e$ for $p$ = 0
was obtained from ARPES spectra  of  the parent Mott insulator Ca$_2$CuO$_2$Cl$_2$
\cite{Ronning98} for states well below the chemical potential. It is an
 approximate value since there is no FS and the usual Fermi liquid mass enhancement
 effects are suppressed.  The limited data  raise the possibility that $m^*/m_e$ could become very large as
$p$ approaches 0.19, the ``special point''  where heat capacity and other measurements
on many hole-doped cuprates suggest that the pseudogap energy scale $E_\mathrm{G}$ goes
to zero. Fig.~4b shows the $p$-dependence of $E_\mathrm{G}$ and  the specific heat jump
at $T_\mathrm{c}$, for YBa$_2$Cu$_3$O$_{6+x}$ \cite{Loram01}. The latter is usually
$\sim \gamma T_\mathrm{c}$ where $\gamma$ is the Sommerfeld coefficient, for example
for a weak coupling BCS superconductor it is equal to $1.43 \gamma T_\mathrm{c}$.

For Y124, every two-dimensional (2D) FS sheet in the BZ will give a contribution to $\gamma$ of $1.46 m^*/m_e
$\,mJ\,mol$^{-1}$K$^{-2}$. This is independent of the number of carriers
in the sheet and arises because in 2D both $\gamma$ and $m^*$ are proportional to the
energy derivative of the FS area \cite{Bergemann03} multiplied by the same enhancement factor.  Our value $m^*=3.0\pm0.3 m_e$ thus
implies a contribution $\gamma=4.4\pm0.4$\,mJ\,mol$^{-1}$K$^{-2}$ for every 2D FS
pocket of the observed frequency present in the BZ. An upper limit obtained from
specific heat measurements of polycrystalline Y124 \cite{Loramunpub}\, is
$\gamma=9$\,mJ\,mol$^{-1}$K$^{-2}$. This is a ``normal state'' value at $T=0$\,K and
zero field, obtained by applying an entropy conserving construction to $\gamma(T)$ from
$T>T_c$ to $T\ll T_c$, and is consistent with the measured jump of
$\delta\gamma=15$\,mJ\,mol$^{-1}$K$^{-2}$ at $T_\mathrm{c}$. If an estimated chain
contribution of 3.5 $\pm 0.5$ \,mJ\,mol$^{-1}$K$^{-2}$ is subtracted, this leaves a
plane contribution $\gamma_\mathrm{plane}$ =  5.5 $\pm 0.5$ \,mJ\,mol$^{-1}$K$^{-2}$.
Hence comparison of heat capacity data with our results casts doubt on the original
model \cite{DoironLeyraud07} involving four hole pockets near the ($\pm\pi/2$,
$\pm\pi/2$) points  where photoemission (ARPES) experiments on underdoped crystals give
evidence for Fermi arcs \cite{Kanigel06}.

Four half-pockets of holes in a reduced BZ still give an electronic heat capacity that is a factor $\sim 8.8/5.5 = 1.6 \pm 0.2$ larger than the above estimate of $\gamma_\mathrm{plane}$. Recent Hall effect measurements
\cite{LeBoeuf07} suggest that the quantum oscillations may be due to a single electron
pocket in the reduced BZ, centered at ($\pi$,0). This would be consistent with
$\gamma_\mathrm{plane}$ but implies that the proposed hole pockets \cite{LeBoeuf07}
only make a very small contribution to the heat capacity.   In contrast to heavy
fermion compounds, where the large heat capacity often suggested that quantum
oscillations from the heavy electrons were not being detected in some of the early
experiments, in the present case it is the small heat capacity that provides
significant constraints to theoretical models for the FS pockets.

The Fermi energy ($E_\mathrm{F}$) can be calculated if we assume that the FS sheets
responsible for the oscillations are nearly 2D, that is, open in the $c$-axis
direction. For a parabolic energy dispersion, we find $E_\mathrm{F}=295$\,K for Y124
and 375\,K for O-II Y123. Intriguingly these are of the same order as the pseudogap
energies $E_\mathrm{G}$ obtained from heat capacity and magnetic susceptibility
\cite{Loram01,Loramunpub} which are $E_\mathrm{G}=570\pm 30$\,K for O-II Y123 and
$E_\mathrm{G}=360\pm 25$\,K for Y124. Note that these values of $E_\mathrm{G}$ are
consistent with the values of $p$ quoted earlier.  If the pockets of carriers are still
present at lower fields and higher $T$,  these low values of $E_\mathrm{F}$ would lead
to $T$-dependent diamagnetism, which although small, would be much more anisotropic
than the spin susceptibility.  This provides another means of testing theoretical
models and making comparisons with ARPES data. Anomalous $T$-dependent magnetic
anisotropy has been detected in the normal state of various cuprate superconductors and
the similarity with Landau-Peierls diamagnetism in the organic conductor HMTSF-TCNQ has
been noted \cite{Miljak}.

In summary, we have observed quantum oscillations in the 80\,K
cuprate superconductor Y124 that have a larger orbit area than in
O-II Y123, with $T_\mathrm{c}$ of 57\,K, and a considerably larger
effective mass. Comparison with heat capacity data places strong constraints
 on the number of pockets present in the BZ, and supports models with a reduced BZ and small FS.

After completing  the present measurements, we became aware of Hall resistivity results
for YBa$_2$Cu$_4$O$_8$ \cite{Bangura07} giving values of $F$ and $m^*$ that agree with
ours.  JRC and EAY thank A. Carrington, S.M. Hayden, N.E. Hussey \cite{Husseynote},
J.W. Loram and J.L. Tallon for helpful discussions and collaboration and the EPSRC
(U.K.) for financial support. This work is supported by DoE grants LDRD-DR-20070085 and
BES Fieldwork grant, ``Science in 100\,T''. Work at NHMFL is performed under the
auspices of the National Science Foundation, DoE and the State of Florida.

\bibliographystyle{aps5etal}

\begin{thebibliography}{28}
\providecommand{\natexlab}[1]{#1}
\providecommand{\bibnamefont}[1]{#1}
\providecommand{\bibfnamefont}[1]{#1}
\providecommand{\citenamefont}[1]{#1}
\providecommand{\url}[1]{\texttt{#1}}
\providecommand{\urlprefix}{URL }
\providecommand{\bibinfo}[2]{#2}
\providecommand{\eprint}[2][]{\url{#2}}

\bibitem[*]{}Present address: School of Physics and Astronomy, University of St Andrews, KY16 9SS, United Kingdom.

\bibitem[{\citenamefont{Doiron-Leyraud} \emph{et~al.}(2007)}]{DoironLeyraud07}
\bibinfo{author}{\bibfnamefont{N.}~\bibnamefont{Doiron-Leyraud}},
  \emph{et~al.}, \bibinfo{journal}{Nature} \textbf{\bibinfo{volume}{447}},
  \bibinfo{pages}{565} (\bibinfo{year}{2007}).


\bibitem[{dop(????)}]{dopingnote}
\bibinfo{note}{The $a$-axis thermopower \cite{Zhou96} suggests
  $p=0.125\pm0.005$ using the relation in Ref.~\cite{Tallon95}. The magnitude
  and $T$-dependence of the $a$-axis resistivity suggest Y124 has nearly the
  same doping as YBa$_2$Cu$_3$O$_{6.75}$ \cite{Segawa01}.}


\bibitem[{EAY(????)}]{EAYthesis}
\bibinfo{note}{E. A. Yelland, PhD Thesis, University of Cambridge, UK (2003)}.


\bibitem[{\citenamefont{Harrison}
  \emph{et~al.}(2007{\natexlab{a}})}]{Harrison07}
\bibinfo{author}{\bibfnamefont{N.}~\bibnamefont{Harrison}}, \emph{et~al.},
  \bibinfo{journal}{Phys. Rev. Lett.} \textbf{\bibinfo{volume}{99}},
  \bibinfo{pages}{056401} (\bibinfo{year}{2007}{\natexlab{a}}).


\bibitem[{\citenamefont{Coffey} \emph{et~al.}(2000)}]{Coffey00}
\bibinfo{author}{\bibfnamefont{T.}~\bibnamefont{Coffey}}, \emph{et~al.},
  \bibinfo{journal}{Rev. Sci. Instr.} \textbf{\bibinfo{volume}{71}},
  \bibinfo{pages}{4600} (\bibinfo{year}{2000}).


\bibitem[{\citenamefont{Mielke} \emph{et~al.}(2001)}]{Mielke01}
\bibinfo{author}{\bibfnamefont{C.}~\bibnamefont{Mielke}}, \emph{et~al.},
  \bibinfo{journal}{J. Phys. Condens. Matter} \textbf{\bibinfo{volume}{13}},
  \bibinfo{pages}{8325} (\bibinfo{year}{2001}).


\bibitem[{res(????)}]{resistivitynote}
\bibinfo{note}{We calculate the amplitude of dHvA oscillations in the
  differential magnetization using the expression for a 2D slab of $k$-space
  given in Fig.~2.11 of [\onlinecite{ShoenbergBook}]. For one orbit of area
  0.024 of the 2D BZ, with $m^*/m_e=3$, a slab thickness $\delta k=2\pi/3c$
  (where $c$ is the spacing between CuO$_2$ bi-layers), a coil filling factor
  of 1/4 and a normal-state $a$-axis resistivity $\rho_a=32\mu\Omega$\,cm, the
  signal amplitude should be 2\,kHz peak-to-peak. For $\rho_a \gtrsim
  32\mu\Omega$\,cm we expect dHvA oscillations to dominate, while for
  $\rho_a\lesssim 32\mu\Omega$\,cm, Shubnikov-de Haas oscillations would do so.
  The $\rho_a$ measured in Ref.~\cite{Bangura07} suggests that we are in the
  regime where both contributions are significant.}


\bibitem[{Sho(????)}]{ShoenbergBook}
\bibinfo{note}{D. Shoenberg, \textit{Magnetic oscillations in metals}
  (Cambridge University Press, Cambridge, UK 1984)}.


\bibitem[{bac(????)}]{backgdnote}
\bibinfo{note}{The subtracted background is of the form
  $c_1+c_2(B-85)+c_3(B-85)^2+10^6/ [1+\exp((B-c_4)/c_5))]$. The first three
  terms describe the nearly linear decrease in $f$ with $B$ which we ascribe to
  the magnetoresistance of the Cu coils; the last term is a non-oscillatory
  phenomenological fit to the high-field tail of the broad step in $f$ centered
  on $B\approx 45$\,T. During some pulses, occasional steps $\sim 1$\,kHz
  occurred in the TDO frequency. An offset was added to $\Delta f$ in some
  field ranges to correct for the discontinuity.}


\bibitem[{\citenamefont{Bascones} \emph{et~al.}(2005)\citenamefont{Bascones,
  Rice, Shorikov, Lukoyanov, and Anisimov}}]{Bascones05}
\bibinfo{author}{\bibfnamefont{E.}~\bibnamefont{Bascones}},
  \bibinfo{author}{\bibfnamefont{T.~M.} \bibnamefont{Rice}},
  \bibinfo{author}{\bibfnamefont{A.~O.} \bibnamefont{Shorikov}},
  \bibinfo{author}{\bibfnamefont{A.~V.} \bibnamefont{Lukoyanov}},
  \bibnamefont{and} \bibinfo{author}{\bibfnamefont{V.~I.}
  \bibnamefont{Anisimov}}, \bibinfo{journal}{Phys. Rev. B}
  \textbf{\bibinfo{volume}{71}}, \bibinfo{pages}{012505}
  (\bibinfo{year}{2005}).


\bibitem[{\citenamefont{Carrington and Yelland}(2007)}]{Carrington07}
\bibinfo{author}{\bibfnamefont{A.}~\bibnamefont{Carrington}} \bibnamefont{and}
  \bibinfo{author}{\bibfnamefont{E.~A.} \bibnamefont{Yelland}},
  \bibinfo{journal}{Phys. Rev. B} \textbf{\bibinfo{volume}{76}},
  \bibinfo{pages}{140508(R)} (\bibinfo{year}{2007}).


\bibitem[{Coo(????)}]{Cooperunpub}
\bibinfo{note}{J. R. Cooper, P. J. Meeson, A. Carrington, L. Balicas, E. A.
  Yelland and B. Dabrowski, 2002, unpublished}.


\bibitem[{Ban(????)}]{Bangura07}
\bibinfo{note}{A. F. Bangura \textit{et al.}, arXiv:cond-mat/07074461}.


\bibitem[{Tno(????)}]{Tnote}
\bibinfo{note}{The values of $\mu_0 H_\mathrm{m}(T)$ have been used to estimate
  potential $T$ errors; the horizontal error bars in Figs.~3(b-c) are based on
  this.}

  \bibitem[{\citenamefont{Segawa and Ando}(2001)}]{Segawa01}
\bibinfo{author}{\bibfnamefont{K.}~\bibnamefont{Segawa}} \bibnamefont{and}
  \bibinfo{author}{\bibfnamefont{Y.}~\bibnamefont{Ando}},
  \bibinfo{journal}{Phys. Rev. Lett.} \textbf{\bibinfo{volume}{86}},
  \bibinfo{pages}{4907} (\bibinfo{year}{2001}).

\bibitem[{Gin(????)}]{GinsbergBook}
\bibinfo{note}{R. M. Hazen, in \textit{Physical Properties of High Temperature
  Superconductors II}, ed. D.M. Ginsberg (World Scientific, Singapore, 1990),
  p. 121}.


\bibitem[{\citenamefont{Zhou} \emph{et~al.}(1996)\citenamefont{Zhou,
  Goodenough, Dabrowski, and Rogacki}}]{Zhou96}
\bibinfo{author}{\bibfnamefont{J.-S.} \bibnamefont{Zhou}},
  \bibinfo{author}{\bibfnamefont{J.~B.} \bibnamefont{Goodenough}},
  \bibinfo{author}{\bibfnamefont{B.}~\bibnamefont{Dabrowski}},
  \bibnamefont{and} \bibinfo{author}{\bibfnamefont{K.}~\bibnamefont{Rogacki}},
  \bibinfo{journal}{Phys. Rev. Lett.} \textbf{\bibinfo{volume}{77}},
  \bibinfo{pages}{4253} (\bibinfo{year}{1996}).


\bibitem[{\citenamefont{Tallon} \emph{et~al.}(1995)\citenamefont{Tallon,
  Bernhard, Shaked, Hitterman, and Jorgensen}}]{Tallon95}
\bibinfo{author}{\bibfnamefont{J.~L.} \bibnamefont{Tallon}},
  \bibinfo{author}{\bibfnamefont{C.}~\bibnamefont{Bernhard}},
  \bibinfo{author}{\bibfnamefont{H.}~\bibnamefont{Shaked}},
  \bibinfo{author}{\bibfnamefont{R.~L.} \bibnamefont{Hitterman}},
  \bibnamefont{and} \bibinfo{author}{\bibfnamefont{J.~D.}
  \bibnamefont{Jorgensen}}, \bibinfo{journal}{Phys. Rev. B}
  \textbf{\bibinfo{volume}{51}}, \bibinfo{pages}{12911} (\bibinfo{year}{1995}).


\bibitem[{Che(????)}]{Chen07}
\bibinfo{note}{W.-Q. Chen, K.-Y. Yang, T. M. Rice and F. C. Zhang,
  arXiv:cond-mat/07063556}.


\bibitem[{\citenamefont{Harrison}
  \emph{et~al.}(2007{\natexlab{b}})\citenamefont{Harrison, McDonald, and
  Singleton}}]{Harrison07b}
\bibinfo{author}{\bibfnamefont{N.}~\bibnamefont{Harrison}},
  \bibinfo{author}{\bibfnamefont{R.~D.} \bibnamefont{McDonald}},
  \bibnamefont{and}
  \bibinfo{author}{\bibfnamefont{J.}~\bibnamefont{Singleton}},
  \bibinfo{journal}{Phys. Rev. Lett.} \textbf{\bibinfo{volume}{99}},
  \bibinfo{pages}{206406} (\bibinfo{year}{2007}{\natexlab{b}}).


\bibitem[{\citenamefont{Trugman}(1990)}]{Trugman90}
\bibinfo{author}{\bibfnamefont{S.~A.} \bibnamefont{Trugman}},
  \bibinfo{journal}{Phys. Rev. Lett.} \textbf{\bibinfo{volume}{65}},
  \bibinfo{pages}{500} (\bibinfo{year}{1990}).


\bibitem[{\citenamefont{LeBoeuf} \emph{et~al.}(2007)}]{LeBoeuf07}
\bibinfo{author}{\bibfnamefont{D.}~\bibnamefont{LeBoeuf}}, \emph{et~al.},
  \bibinfo{journal}{Nature} \textbf{\bibinfo{volume}{450}},
  \bibinfo{pages}{533} (\bibinfo{year}{2007}).


\bibitem[{\citenamefont{Ronning} \emph{et~al.}(1998)}]{Ronning98}
\bibinfo{author}{\bibfnamefont{F.}~\bibnamefont{Ronning}}, \emph{et~al.},
  \bibinfo{journal}{Science} \textbf{\bibinfo{volume}{282}},
  \bibinfo{pages}{2067} (\bibinfo{year}{1998}).


\bibitem[{\citenamefont{Loram} \emph{et~al.}(2001)\citenamefont{Loram, Luo,
  Cooper, Liang, and Tallon}}]{Loram01}
\bibinfo{author}{\bibfnamefont{J.~W.} \bibnamefont{Loram}},
  \bibinfo{author}{\bibfnamefont{J.}~\bibnamefont{Luo}},
  \bibinfo{author}{\bibfnamefont{J.~R.} \bibnamefont{Cooper}},
  \bibinfo{author}{\bibfnamefont{W.~Y.} \bibnamefont{Liang}}, \bibnamefont{and}
  \bibinfo{author}{\bibfnamefont{J.~L.} \bibnamefont{Tallon}},
  \bibinfo{journal}{J. Phys. Chem. Solids} \textbf{\bibinfo{volume}{62}},
  \bibinfo{pages}{59} (\bibinfo{year}{2001}).


\bibitem[{Lor(????)}]{Loramunpub}
\bibinfo{note}{J. W. Loram, APS March Meeting, Denver, Colorado, 2007
  unpublished}.


\bibitem[{\citenamefont{Bergemann} \emph{et~al.}(2003)\citenamefont{Bergemann,
  Mackenzie, Julian, Forsythe, and Ohmichi}}]{Bergemann03}
\bibinfo{author}{\bibfnamefont{C.}~\bibnamefont{Bergemann}},
  \bibinfo{author}{\bibfnamefont{A.~P.} \bibnamefont{Mackenzie}},
  \bibinfo{author}{\bibfnamefont{S.~R.} \bibnamefont{Julian}},
  \bibinfo{author}{\bibfnamefont{D.}~\bibnamefont{Forsythe}}, \bibnamefont{and}
  \bibinfo{author}{\bibfnamefont{E.}~\bibnamefont{Ohmichi}},
  \bibinfo{journal}{Adv. Phys.} \textbf{\bibinfo{volume}{52}},
  \bibinfo{pages}{639} (\bibinfo{year}{2003}).


\bibitem[{\citenamefont{Kanigel} \emph{et~al.}(2006)}]{Kanigel06}
\bibinfo{author}{\bibfnamefont{A.}~\bibnamefont{Kanigel}}, \emph{et~al.},
  \bibinfo{journal}{Nature Physics} \textbf{\bibinfo{volume}{2}},
  \bibinfo{pages}{447} (\bibinfo{year}{2006}).


\bibitem[{\citenamefont{Miljak} \emph{et~al.}(1990)}]{Miljak}
\bibinfo{author}{\bibfnamefont{M.}~\bibnamefont{Miljak}}, \emph{et~al.},
  \bibinfo{journal}{Phys. Rev. B} \textbf{\bibinfo{volume}{42}},
  \bibinfo{pages}{10742} (\bibinfo{year}{1990}).


\bibitem[{Hus(????)}]{Husseynote}
\bibinfo{note}{N. E. Hussey showed both of us preliminary evidence for quantum
  oscillations in the Hall coefficient of one Y124 crystal (C. Proust
  \textit{et al.}, unpublished) two weeks before the pulsed field experiments
  reported here.}


\end{thebibliography}

\end{document}